\documentclass[aps,pra,twocolumn,floatfix,superscriptaddress]{revtex4-2}

\usepackage{bm,bbm,amsmath, amssymb,amsbsy, amsthm,graphicx,subfigure,color,txfonts,multirow,enumitem} 
\usepackage[framemethod=tikz]{mdframed}
\usepackage[export]{adjustbox}
\usepackage[colorlinks]{hyperref}
\hypersetup{
	bookmarksnumbered,
	pdfstartview={FitH},
	citecolor={blue},
	linkcolor={blue},
	urlcolor={blue},
	pdfpagemode={UseOutlines}
}
\usepackage{cleveref}

\newcommand{\ket}[1]{|#1\rangle}
\newcommand{\bra}[1]{\langle#1|}

\newcommand{\tr}[1]{\mathrm{tr}\left\{#1\right\}}

\newcommand*\sq{\mathbin{\vcenter{\hbox{\rule{.3ex}{.3ex}}}}}

\begin{document}
 
 \title{Quantitative bounds to propagation of  quantum correlations in many-body systems}
 


 \author{Davide Girolami}
\email{davegirolami@gmail.com}
 \affiliation{$\hbox{DISAT, Politecnico di Torino, Corso Duca degli Abruzzi 24, Torino 10129, Italy}$
 }
\author{Michele Minervini} 
\affiliation{$\hbox{DISAT, Politecnico di Torino, Corso Duca degli Abruzzi 24, Torino 10129, Italy}$
 }
\begin{abstract} 
We  investigate how much information about a quantum system can be simultaneously communicated to independent observers, by establishing quantitative limits to bipartite quantum correlations in many-body systems. As recently reported in Phys. Rev. Lett. 129, 010401 (2022), bounds on quantum discord and entanglement of formation between a single quantum system and its environment, e.g., a large number of photons, dictate that independent observers which monitor environment fragments inevitably acquire only classical information about the system. Here, we corroborate and generalize those findings. First, we calculate continuity bounds of quantum discord,  which establish how much states with a small amount of quantum correlations deviate from being embeddings of classical probability distributions.  Also, we demonstrate a  universally valid upper bound to the bipartite entanglement of formation between an arbitrary pair of components of a many-body quantum system. The results confirm that proliferation of classical information in the Universe suppresses quantum correlations.
\end{abstract}

\date{\today}

\maketitle

\section*{Introduction.} 
 
\noindent Quantum systems display correlations that cannot be explained by the laws of classical probability \cite{EPR,epr2,entanglement}. Such a counterintuitive feature of the quantum world  signals a dramatic departure from what we perceive to be our macroscopic reality.  Also, quantum correlations promise to be the key resources for quantum technologies, as they allow   to overperform classical devices in computing, communication, and sensing \cite{telep,super,capa,naturedisc,tech}. Indeed, terms like ``Entanglement'' are becoming common parlance in many branches of science. \\

\noindent The co-existence between classical and quantum regimes in our Universe, and for all practical purposes between our laptops and future quantum computers, is justified {\it within} quantum theory by means of bounds to quantum communication.  We are free to share {\it classical} information, i.e., the outcome of a measurement on a physical system,  with an arbitrary number of observers. That is, Nature allows to copy  bits of information and simultaneously distribute them to many independent receivers, which then reach an agreement about the measured quantity.  As a result, a defining feature of our description of the world is that properties of physical systems acquire the status of  ``objective''.\\
 \noindent Yet, fundamental results like the no-cloning theorem \cite{noclone}, and monogamy relations of entanglement measures \cite{mono}, set  limits to broadcasting {\it quantum} information, i.e., the wavefunction of a quantum system. Further recent works have demonstrated constraints to the concurrent distribution of quantum information from a single source to a network of observers, formalized in terms of bounds to quantum correlations \cite{red,brandao,ranard,adesso,akram,akram2}. Their operational meaning is that the very quantum theory dictates that quantum information cannot be concurrently stored and made available to independent observers. Consequently, these agents would never reach consensus on  quantum properties of the source. \\

\noindent These results support the core ideas underpinning Quantum Darwinism, a genuinely quantum explanation of the emergence of a classical macroscopic reality \cite{darwin1,darwin2,darwin4,darwin5,darwin6,exp2,pan,ollivier2,korbiczreview}. Interactions between physical systems and their environment select pointer states \cite{deco}, which encode effectively classical  information that is copied and redundantly spread into the environment. They are the only kind of knowledge that can be concurrently
 acquired   by many  observers. 
The reason is that  such non-cooperating observers  obtain information about a system by eavesdropping  on small, distinct fractions of the system environment, i.e., scattered photons \cite{Z07,Z13,Zurek81,deco}. \\

\noindent In this paper, we review and extend the findings of Ref. \cite{red}. As a preliminary step,  we recall quantitative bounds to the  quantum discord \cite{discordzurek}, the most general kind of quantum correlation, and the entanglement of formation \cite{entform}, between a system of interest and fragments of its environment.  In particular, we show the emergence of the bound to the entanglement of formation with a numerical study of the correlation pattern in a star-like quantum network. \\
These bounds are universally valid (they hold for any global pure state of the system and the environment), confirming that quantum Darwinism is a generic feature of many-body quantum systems \cite{brandao}. Further, they are easy to compute: this is surprising, since the quantification of quantum correlations in complex, multipartite systems is generally a hard problem  \cite{tech,corr1,corr2,discorev,sanpera,newzurek,sen,discome}, and neither quantum discord nor the entanglement of formation are monogamous   \cite{koashi,mono,braga,monogamy,acin}.  Moreover, these upper limits are physically meaningful: they are expressed in terms of measures of (dis)-agreement among  observers that eavesdrop on the environment about the received information, which is inevitably classical.  That is, whenever we reach consensus, a prominent feature of classical reality,  quantum information is inaccessible.   
Only a utopian observer able to intercept  large fractions of the environment (specifically, more than half of the scattered photons carrying relevant information  \cite{red})
could establish non-negligible quantum correlations with the system under scrutiny. \\

\noindent Then, we present and demonstrate other quantitative constraints to quantum correlations in many-body systems. We firstly focus on quantum discord. We prove continuity bounds nearby the set of states which describe classically correlated systems. 
In particular, by employing the relative entropy as (pseudo)-distance,  we demonstrate that quantum discord takes small values for density matrices that are close to the set of ``classical-quantum'' states \cite{tech}. The result implies that quantum information cannot be communicated via interactions that are described, with arbitrarily small error, by classical physics. A spectacular example is, in fact, the measurements performed by we humans on macroscopic objects. \\
Second, we generalize the upper bound to the bipartite entanglement of formation. By introducing a new measure of (dis)-agreement among observers about classical information, we derive a limit to the average entanglement that an arbitrary component of a many-particle quantum system can share with other parts. The larger the system,  the smaller is the amount of entanglement that can be locally established. Therefore, bipartite quantum correlations are suppressed, even if the global state displays genuine multipartite entanglement. \\

\noindent The paper is organized as follows. In Section \ref{sec1}, we introduce the information-theoretic measures of classical and quantum correlations that we will employ here. In Section \ref{sec2},  we review the main results of Ref.~\cite{red}. In Section \ref{sec3}, we demonstrate  the continuity bounds to quantum discord and a generalized bound to the bipartite entanglement of formation in many-body systems. In the Conclusion, we will outline our findings and suggest open questions that are worthy of investigation.

\section{Measures of classical and quantum correlations}\label{sec1} 

\begin{figure}
\includegraphics[width=.46\textwidth,height=5.5cm,,fbox=0.5pt 5pt]{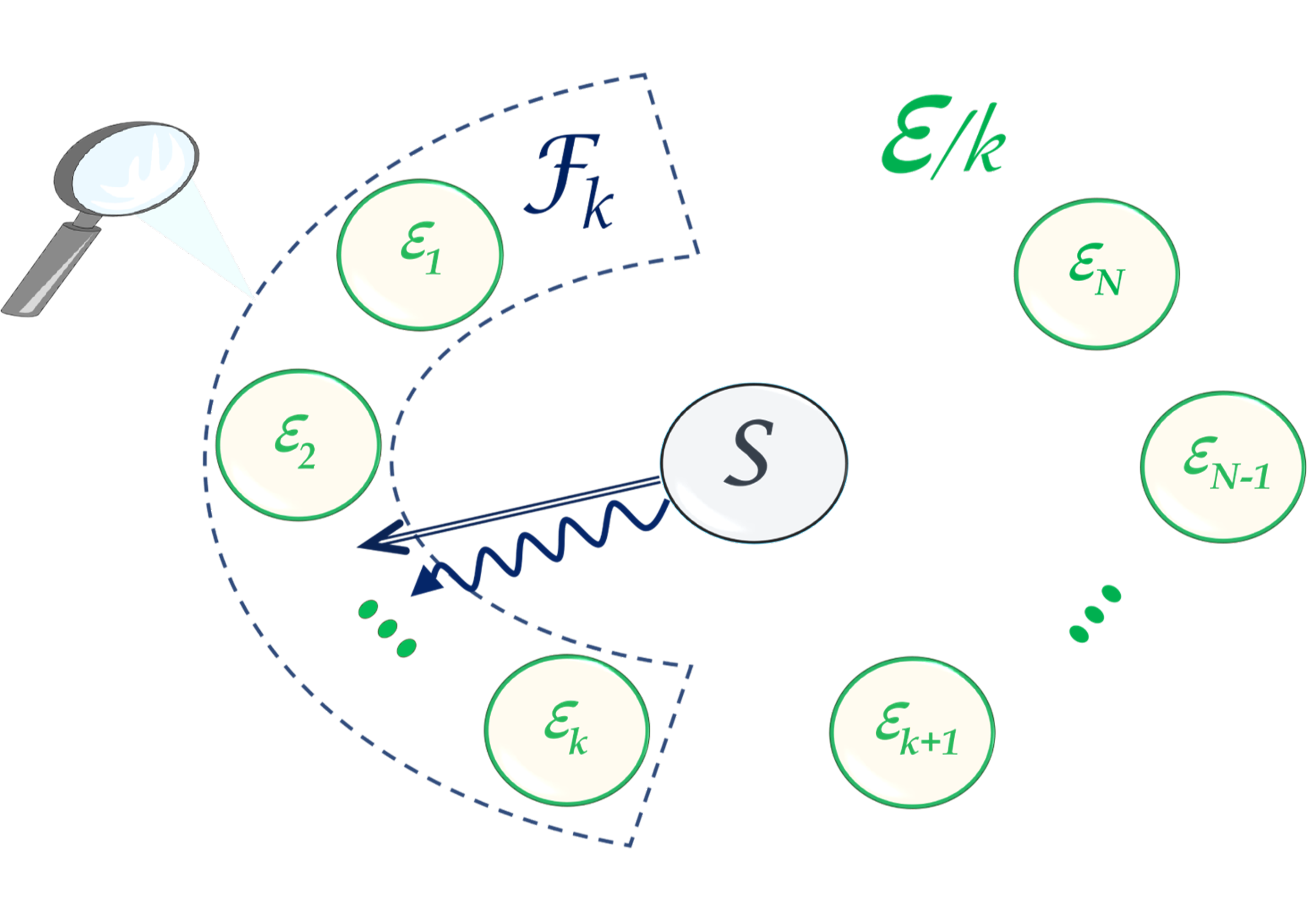}
\caption{We consider a quantum Universe in which a system ${\cal S}$ interacts with an $N$-particle environment ${\cal E}$. We investigate fundamental bounds to bipartite quantum correlations, as quantified by quantum discord and the entanglement of formation, between ${\cal S}$ and an environment fragment ${\cal F}_k$.}
\label{fig3}
\end{figure}

\noindent Consider a  quantum Universe that consists of a quantum system ${\cal S}$ and its $N$-partite environment  ${\cal E}:=\cup_{i=1}^N\varepsilon_i$  (FIG.~\ref{fig3}).  We define an environment fragment of  $k\leq N$ particles  ${\cal F}_{k}:=\cup_{\# i=k}\varepsilon_i$ and its complement ${\cal E}_{/k}:={\cal E}/{\cal F}_k$.   In the following, we recall the definitions of widely employed measures of classical and quantum correlations between ${\cal S}$ and the fragment ${\cal F}_k$. 

\noindent Being $H(\rho_{{\cal X}}):=-\tr{\rho_{\cal X}\log_2 \rho_{\cal X}}$  the von Neumann entropy of the state $\rho_{{\cal X}}$ of the system ${\cal X}$, the statistical dependence between ${\cal S}$ and  ${\cal F}_{k}$ in the state $\rho_{{\cal S}{\cal F}_{k}}$ is given by the mutual information 
\begin{eqnarray}
I(\rho_{{\cal S}{\cal F}_{k}}):=H(\rho_{{\cal S}})+H(\rho_{{\cal F}_{k}})-H(\rho_{{\cal S}{\cal F}_{k}}).
\end{eqnarray}
 The quantity evaluates the total  information shared by two systems. Remarkably, it splits into classical and quantum components \cite{discordzurek,vedral}.  \\
 The classical part is constructed as follows. Suppose one  performs a local  positive operator-valued measure (POVM)  $\mathbf{M}_k:=\left\{\mathbf{M}_\alpha, \sum_\alpha \mathbf{M}_\alpha^\dagger\mathbf{M}_\alpha= \mathbb{I}\right\}$ on ${\cal F}_{k}$. The post-measurement state of the bipartition ${\cal SF}_k$ is
 \begin{equation}
 \rho_{{\cal SF}_{k,\mathbf{M}_k}}=\sum_\alpha \left(\mathbb{I}\otimes \mathbf{M}_{\alpha}\right)\,\rho_{{\cal SF}_k} \left(\mathbb{I}\otimes \mathbf{M}_\alpha^{\dagger}\right).
 \end{equation}
 Then, classical correlations are quantified as the maximal information about ${\cal S}$ an observer can extract  by measurements on ${\cal F}_k$   \cite{vedral,holevo}, which  is given by the maximal mutual information of the post-measurement state:
\begin{eqnarray}\label{class}
J\left(\rho_{{\cal S}\check{\cal F}_{k}}\right):=\max\limits_{\mathbf{M}_k} I\left(\rho_{{\cal S}{\cal F}_{k,\mathbf{M}_k}}\right).
\end{eqnarray}
The  maximal value of classical correlations, i.e., the maximal classical information that can flow from ${\cal S}$ to an environment fragment, is $H(\rho_{{\cal S}})$. \\
The quantum part of the mutual information, namely  \emph{quantum discord}, is then defined as the difference between pre- and post-measurement mutual information \cite{discordzurek}:
 \begin{equation}
 D\left(\rho_{{\cal S}\check{\cal F}_{k}}\right):=I\left(\rho_{{\cal S}{\cal F}_{k}}\right)-J\left(\rho_{{\cal S}\check{\cal F}_{k}}\right).
 \end{equation}
This quantity is the minimal {\it quantum} information about ${\cal S}$  that  is lost by ${\cal F}_k$  when it is  subject to a local measurement $\mathbf{M}_{k}$  \cite{streltsovzurek,wiseman,EPR}.\\
 \noindent Quantum discord has captured a lot of interest  because of its peculiar properties. It can exist without entanglement and, conversely to entanglement, can be created by local operations and classical communication (LOCCs). Therefore, it has been considered for some time an appealing alternative to entanglement as  a resource for quantum information processing  \cite{discordzurek,acin,locc,convert,StreltsovBruss,adessopiani,lqu}.\\
\noindent  Note that, for pure states of ${\cal SF}_k$, quantum discord is  equal to the entanglement entropy: $ D\left(\rho_{{\cal S}\check{\cal F}_{k}}\right)= D\left(\rho_{\check{\cal S}{\cal F}_{k}}\right)=H(\rho_{{\cal S}})$, reaching the maximal value  $H(\rho_{{\cal F}_k})$ for maximally entangled states \cite{tech}.   Yet, for mixed states, classical and quantum correlations are in general not invariant under permutations of a bipartition components: $J\left(\rho_{{\cal S}\check{\cal F}_{k}}\right)\neq J\left(\rho_{\check{\cal S}{\cal F}_{k}}\right)$, and $D\left(\rho_{{\cal S}\check{\cal F}_{k}}\right)\neq D\left(\rho_{\check{\cal S}{\cal F}_{k}}\right)$. Further, $D\left(\rho_{{\cal S}\check{\cal F}_k}\right)=0$ does not imply $D\left(\rho_{\check{\cal S}{\cal F}_k}\right)=0$. 
 
 \noindent Next, we review the definition of entanglement of formation  of a state $\rho_{{\cal SF}_k}=\sum_\alpha p_\alpha\rho_{\alpha,{\cal SF}_k}$ \cite{entform}, with $\rho_{\alpha}=\ket{\alpha}\bra{\alpha}$. It is obtained by convex roof optimization of the entanglement entropy:
  \begin{eqnarray}
  E(\rho_{{\cal S}{\cal F}_k}):=\min\limits_{\left\{p_\alpha,\rho_\alpha\right\}} -\sum_\alpha p_\alpha \text{tr}\left\{ \text{tr}_{{\cal S}}\left\{\rho_{\alpha,{\cal SF}_k}\right\}\log_2\text{tr}_{{\cal S}}\left\{\rho_{\alpha,{\cal SF}_k}\right\}\right\}.
  \end{eqnarray}
 There exists a surprising trade-off relation between the  entanglement of formation and classical correlations in tripartite systems, discovered by Koashi and Winter \cite{koashi}: 
\begin{equation}\label{koa}
E(\rho_{{\cal S}{\cal F}_k})\leq\, H(\rho_{{\cal S}})-J\left(\rho_{{\cal S}\check{\cal E}_{/k}}\right).
\end{equation}
 The inequality  is saturated for pure states of the Universe ${\cal SE}$. There is no loss of generality in this assumption: every mixed state of the Universe can be purified by dilation. The result has been employed to derive quantitative relations between quantum discord and entanglement of formation \cite{fanchini}.

\section{Quantitative bounds to bipartite quantum correlations in many-body systems}\label{sec2}
\noindent In this Section, we review the main results of Ref.~\cite{red}. We focus on setting bounds to correlations between ${\cal S}$ and single-site subsystems $\varepsilon_i$. The results apply to  fragments  of arbitrary size ${\cal F}_k$ straightforwardly. \\Classical information about ${\cal S}$ can be freely cloned and simultaneously distributed to the environment fragments.
For instance, consider creation of classical correlations  in a three-bit register by a  XOR gate, $\ket{0}\bra{0}_{{\cal F}_k}(\ket{00}\bra{00}+\ket{11}\bra{11})/2_{{\cal SE}_{/k}}\rightarrow(\ket{000}\bra{000}+\ket{111}\bra{111})/2_{{\cal F}_k{\cal SE}_{/k}}$. More generally, one can saturate the inequality $\bar{J}\left(\rho_{{\cal S}\check\varepsilon_{i}}\right):= \frac1N\sum_{i=1}^N  J\left(\rho_{{\cal S}\check\varepsilon_{i}}\right) \leq H(\rho_{{\cal S}})$. \\
\begin{figure}
\includegraphics[width=.46\textwidth,height=4.5cm,fbox=0.5pt 5pt]{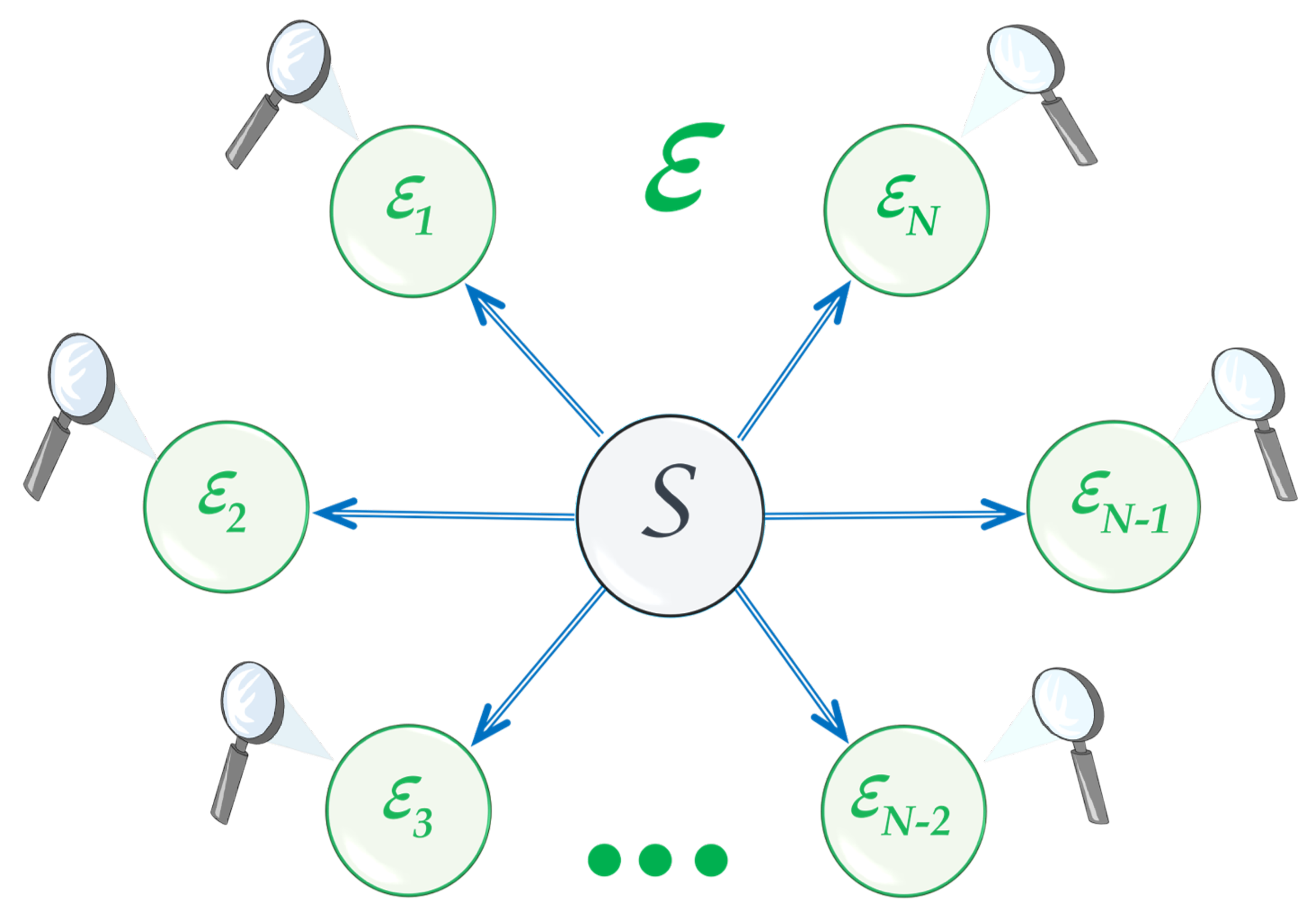}
\caption{There exist upper limits to bipartite quantum correlations  between a system  ${\cal S}$ and the environment subsystems $\varepsilon_i$s of ${\cal E}$. Proliferation of classical information, as quantified by the amount of consensus about ${\cal S}$ that can be reached by observers eavesdropping on $\varepsilon_i$s, inhibits quantum discord and entanglement.}
\label{fig1}
\end{figure}
\noindent Conversely, quantum correlations are restricted by the very same quantum laws. The inner mechanism suppressing quantum information is the creation of consensus among a large number of observers that access copies of classical information deposited in different environment fragments (FIG.~\ref{fig1}). Let us quantify the average (dis)-agreement about the classical information on ${\cal S}$ that an observer tracking  a particle $\varepsilon_i$ experiences with another agent that accesses the rest of the environment ${\cal E}_{/i}$: one defines the parameters
\begin{align}\label{delta}
\delta:=&\sum_{i=1}^{N} \delta_i/N,\\
\delta_i:=&\ \frac{
J\left(\rho_{{\cal S}\check{{\cal E}}}\right)-\min\left\{J\left(\rho_{{\cal S}\check{\varepsilon}_i}\right),J\left(\rho_{{\cal S}\check{\cal E}_{/i}}\right)\right\}}{H(\rho_{{\cal S}})}\, \in\,[0,1].\nonumber
\end{align}

 \begin{figure*}
\subfigure{\fbox{\hspace{25pt}\includegraphics[width=.4\textwidth,height=4.3cm]{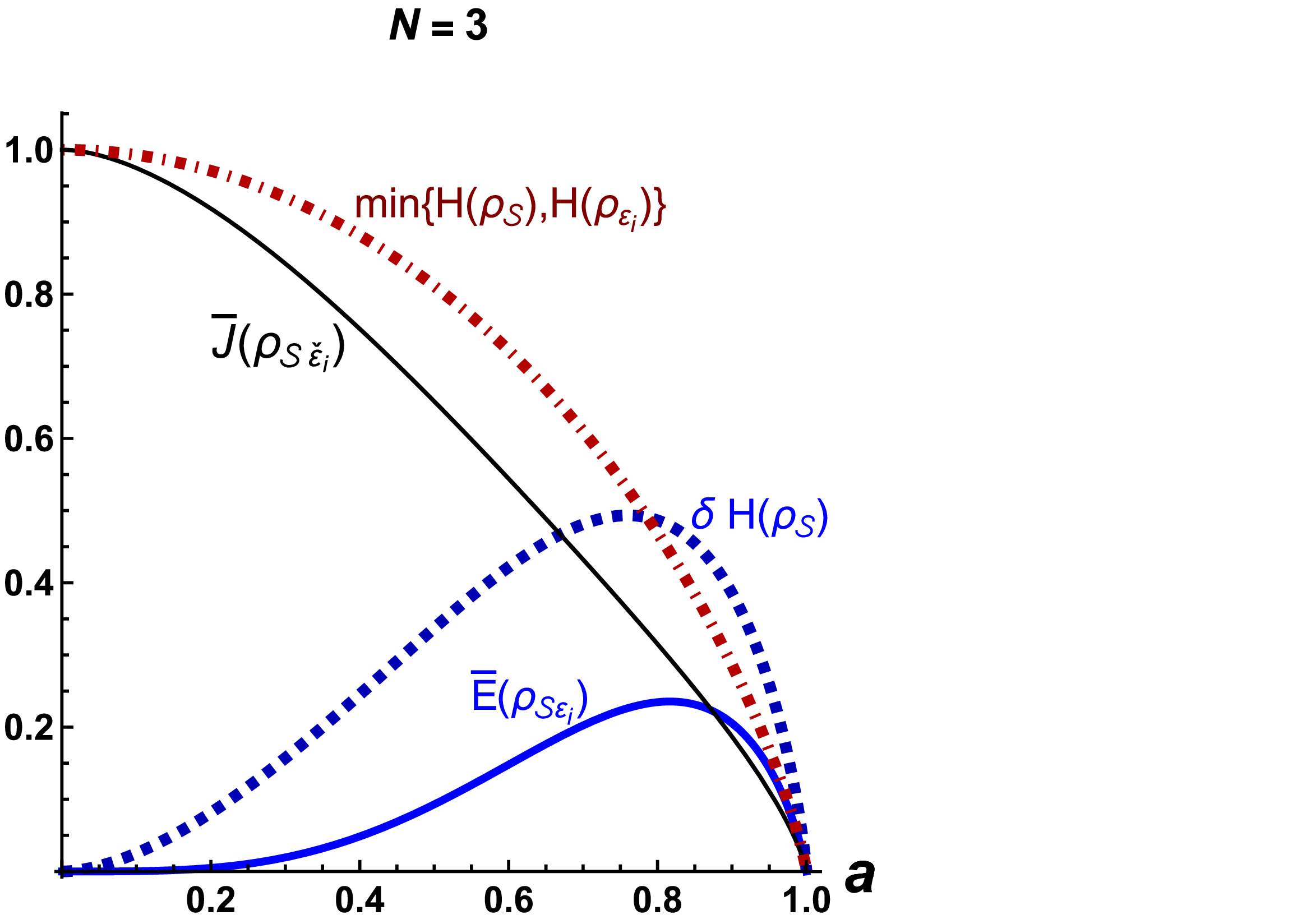}\hspace{-50pt}
\includegraphics[width=.4\textwidth,height=4.3cm]{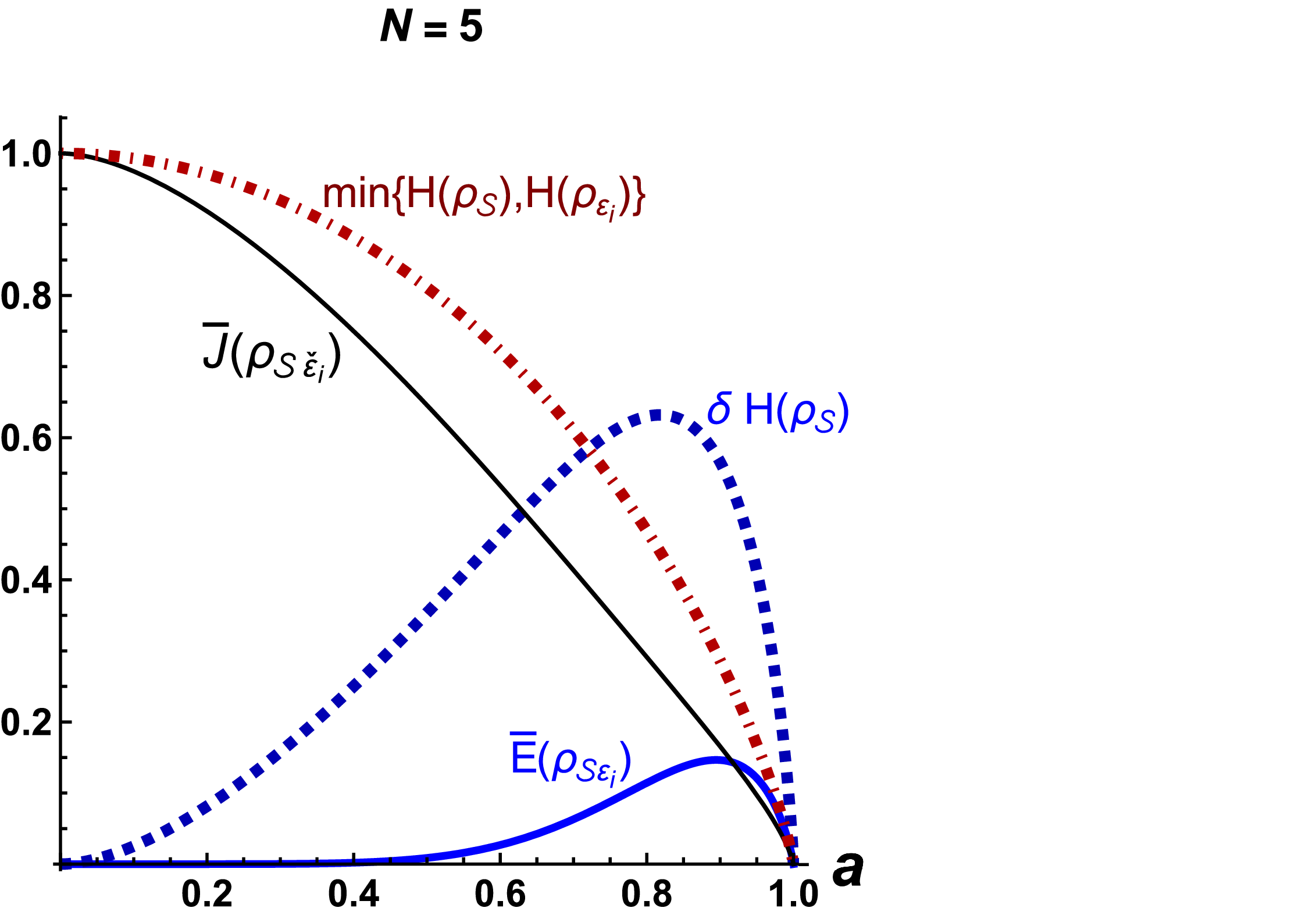}\hspace{-50pt}
\includegraphics[width=.4\textwidth,height=4.3cm]{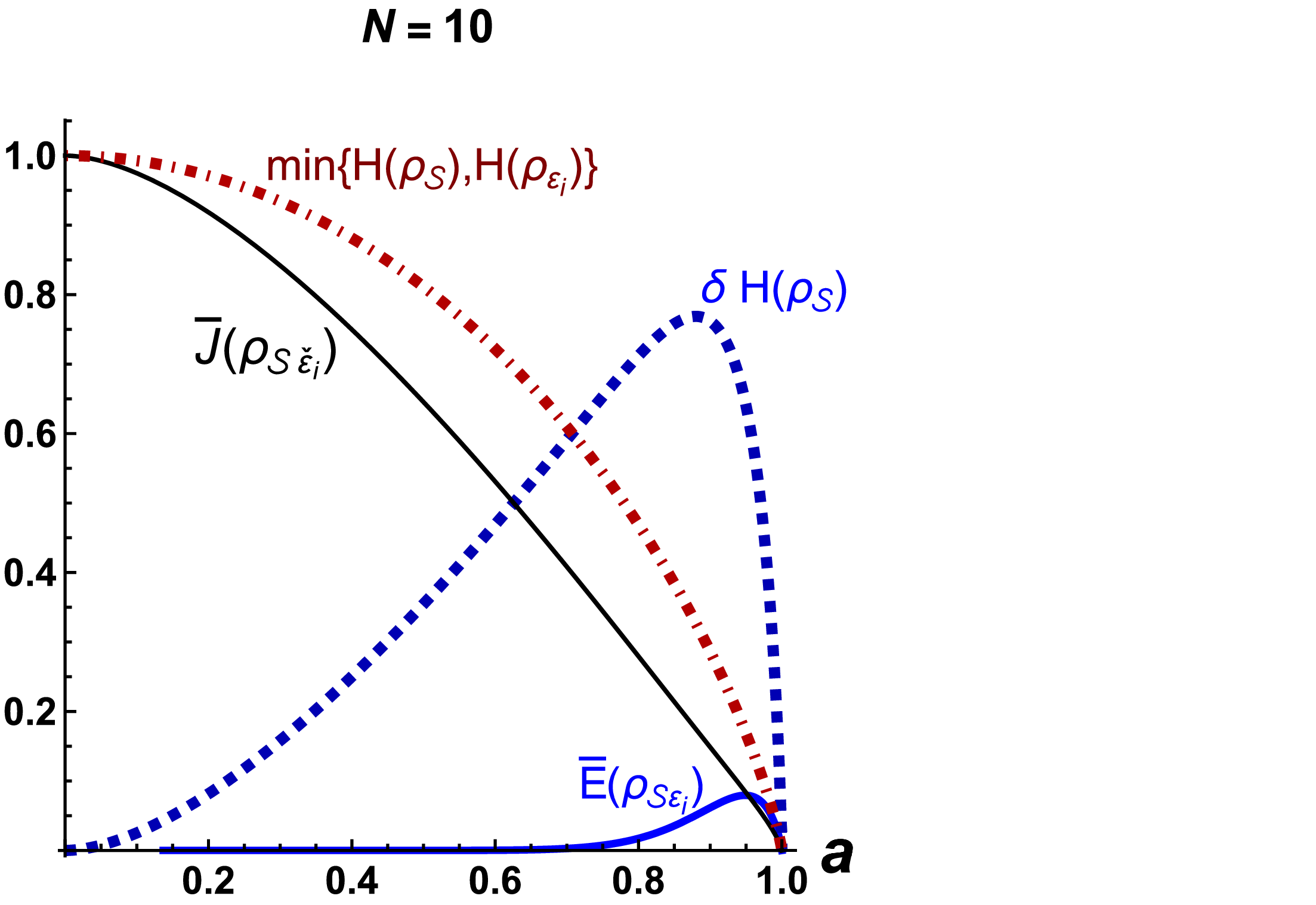}\hspace{-40pt}}}
\caption{
We show the bound to the average entanglement of formation (Eq.~(\ref{entbound})) in action. The following quantities are computed in the final state ${\bf U}_{{\cal SE}}(a)\ket{+}_{\cal S}\ket{0}_{\cal E}^{\otimes N}$, for different values of $N$ (see the main text for full details): the average entanglement of formation, $\bar E\left(\rho_{{\cal S}\varepsilon_{i}}\right)$ (blue line \textcolor{rgb:red,0;green,0;blue,0.5}{{\large ---}});  the upper bound in Eq.~(\ref{entbound})  (dashed blue line \textcolor{rgb:red,0;green,0;blue,0.5}{{\large$\sq\sq\sq$}}); the known entropic upper bound $\min\left\{H(\rho_{{\cal S}}),H(\rho_{\varepsilon_{i}})\right\}$ (dotdashed bordeaux line \textcolor{rgb:red,0.3;green,0;blue,0}{{\large$\sq\cdot\sq\cdot\sq$}}); the  average classical correlations $\bar J\left(\rho_{{\cal S}\check{\varepsilon}_{i}}\right)$ (black line \textcolor{black}{{\large ---}}).  The average values of classical and quantum correlations are the ones computed for an arbitrary pair ${\cal S}\varepsilon_i$, because of the symmetry under environment subsystem permutations of the final state of the Universe. The newfound bound is much more informative than the known entropic bound in the limit $a\rightarrow 0$, when the global state of the Universe  is highly entangled. The study confirms that proliferation of a larger amount of classical correlations suppresses quantum correlations.  Comparing these results with Fig.~3 of Ref.\cite{red}, we note that the entanglement of formation declines by increasing  $N$ much faster than quantum discord.}
\label{fig2}
\end{figure*}
 \noindent We briefly discuss why the index $\delta_i$ is a good measure of the  (lack of) consensus between two observers monitoring $\varepsilon_i$ and ${\cal E}_{/i}$, respectively.  Assume $J\left(\rho_{{\cal S}\check{{\cal E}}_{/i}}\right)\geq J\left(\rho_{{\cal S}\check{\varepsilon}_{i}}\right)$. If $\delta_i=0$, then $J\left(\rho_{{\cal S}\check{{\cal E}}
}\right)=J\left(\rho_{{\cal S}\check{{\cal E}}_{/i}}\right)=J\left(\rho_{{\cal S}\check{\varepsilon}_{i}}\right)$. The reverse implication is also true.  Hence, the parameter $\delta_i$ is zero if and only if  the same classical information about ${\cal S}$ is simultaneously available into $\varepsilon_i$ and ${\cal E}_{/i}$. That is, if and only if observers measuring on the two environment fragments are in perfect agreement. Further, if $\delta_i=1$, then $J\left(\rho_{{\cal S}\check{\varepsilon}_i}\right)=0$, and there is maximal disagreement between the observers.  The reverse statement holds too. 
 
\noindent Introducing a measure of (lack of) objectivity about classical information is instrumental in proving a bound to bipartite quantum discord in many-body systems for any pure state of the universe $\ket{\psi}_{{\cal SE}}$. \\
The result can be demonstrated as follows. First, one observes that
 \begin{align*}
  I(\rho_{{\cal S}\varepsilon_{i}})+ I(\rho_{{\cal S}{\cal E}_{/i}})&= 2\, H(\rho_{{\cal S}}),\,\forall\,i\Rightarrow\\
\nonumber
 I(\rho_{{\cal S}\varepsilon_{i}})+ I(\rho_{{\cal S}{\cal F}_{k}}) &\leq 2\,H(\rho_{{\cal S}}),\,\forall\, i,\,\forall\, {\cal F}_k\subseteq {\cal E}_{/i}\Rightarrow\nonumber\\
\bar I(\rho_{{\cal S}{\cal E}_{/i}}):&=\frac1N  \sum_{i=1}^N  I(\rho_{{\cal S}{\cal E}_{/i}})\geq H(\rho_{{\cal S}}).
\end{align*}  
This preliminary result implies an upper bound to quantum discord when the state of the Universe is pure $\left(H(\rho_{{\cal S}})=J\left(\rho_{{\cal S}\check{\varepsilon}_i}\right)\right)$:
\begin{align*}
 I(\rho_{{\cal S}\varepsilon_{i}})&= 2\, H(\rho_{{\cal S}})- I(\rho_{{\cal S}{\cal E}_{/i}})\Rightarrow\\
 D(\rho_{{\cal S}\check{\varepsilon}_i})&= 2\, H(\rho_{{\cal S}})-  J(\rho_{{\cal S}\check{\varepsilon}_i})-I(\rho_{{\cal S}{\cal E}_{/i}})\\
  D(\rho_{{\cal S}\check{\varepsilon}_i})&\leq 2\, H(\rho_{{\cal S}})+(\delta_i-1)\, H(\rho_{{\cal S}})-I(\rho_{{\cal S}{\cal E}_{/i}})\\
  &\leq (\delta_i+1)\,H(\rho_{{\cal S}})-I(\rho_{{\cal S}{\cal E}_{/i}}).
 \end{align*}
Therefore, there exists a related bound to the average quantum discord between central system and environment subsystem:
\begin{align}
  \label{main}
 \bar{D}(\rho_{{\cal S}\check{\varepsilon}_i}):&= \frac1N\sum_{i=1}^N D(\rho_{{\cal S}\check{\varepsilon}_i}),\nonumber\\
  \bar{D}(\rho_{{\cal S}\check{\varepsilon}_i})& \leq   \frac1N\sum_{i=1}^N\left\{(\delta_i +1)\,H(\rho_{{\cal S}})-I(\rho_{{\cal S}{\cal E}_{/i}})\right\}\nonumber\\
  &\leq \, (\delta +1)\,H(\rho_{{\cal S}})-\bar I(\rho_{{\cal S}{\cal E}_{/i}})\Rightarrow \nonumber\\
  \bar{D}\left(\rho_{{\cal S}\check\varepsilon_{i}}\right)&\leq\,
\delta\,H(\rho_{{\cal S}}).
 \end{align}
 
\noindent Hence, consensus about classical information, i.e., the emergence of classical objectivity about properties of ${\cal S}$ by indirect observation (intercepting fragments of the environment), suppresses quantum correlations.\\
An equivalent bound holds for the entanglement of formation. By employing the Koashi-Winter inequality in Eq.~(\ref{koa}), and reminding the definition in Eq.~(\ref{delta}), a few algebra steps show  that, for pure states of the universe, one has
\begin{align}\label{entbound}
E(\rho_{{\cal S}\varepsilon_i})=&\, H(\rho_{{\cal S}})-J\left(\rho_{{\cal S}\check{\cal E}_{/i}}\right)\nonumber \\
\leq& \, H(\rho_{{\cal S}}) -\min\left\{J\left(\rho_{{\cal S}\check{\varepsilon}_i}\right),J\left(\rho_{{\cal S}\check{\cal E}_{/i}}\right)\right\}\nonumber\\ 
\leq&\, \delta_i\,H(\rho_{{\cal S}})\Rightarrow\nonumber\\
\bar{E}\left(\rho_{{\cal S}\varepsilon_{i}}\right):=&\frac1N\sum_{i=1}^N  E\left(\rho_{{\cal S}\varepsilon_{i}}\right)\leq\,
\delta\,H(\rho_{{\cal S}}).
\end{align}

\noindent We elucidate the meaning of the bound with a numerical study. Consider the quantum Universe to be in the initial uncorrelated state $\ket{+}_{{\cal S}} \ket{0}^{\otimes N}_{{\cal E}}$. Then, one applies the  unitary  ${\bf U}_{{\cal SE}}(a)\equiv\Pi_{i=1}^N {\bf U}_{{\cal S}\varepsilon_i}(a)$, where the two-site transformation ${\bf U}_{{\cal S}\varepsilon_i}(a)$ is the ``c-maybe'' gate on ${\cal S}\varepsilon_i$ \cite{akram}:
\[\left(\begin{array}{cccc}

1 & 0 & 0 & 0 \\
0 & 1 & 0 & 0 \\
0 & 0 & a & \sqrt{1-a^2} \\
0 & 0 & \sqrt{1-a^2} & -a 
\end{array}\right), \,\,a\in[0,1]. 
\]
 This algorithm models the interaction of a quantum system ${\cal S}$ with a large photonic environment \cite{photon2,darwin2}. \\
 We calculate the local average  bipartite classical correlations $\bar J\left(\rho_{{\cal S}\check{\varepsilon}_{i}}\right)$ and the average entanglement of formation $\bar E\left(\rho_{{\cal S}\varepsilon_{i}}\right)$ in the final state ${\bf U}_{{\cal SE}}(a)\ket{+}_{\cal S}\ket{0}_{\cal E}^{\otimes N}$.  Their values can be computed analytically \cite{ranktwo,akram}.  Also, we calculate the newfound bound   Eq.~(\ref{entbound})  and compare it against the known upper limit to the entanglement of formation, given by the smallest between the marginal entropies $H(\rho_{{\cal S}}), H(\rho_{\varepsilon_i})$. \\
The results are plotted in FIG.~\ref{fig2}. They show how the entanglement of formation obeys a ``weak monogamy relation'' dictated by the abundance of classical information about ${\cal S}$ simultaneously available throughout the environment. For  $a\rightarrow 0$, the universe comes close to be in a (generalized) GHZ state and such a behaviour is  magnified:  quantum correlations in partitions ${\cal S}\varepsilon_i$ vanish, while classical information proliferation is maximized. \\
This simple yet instructive model also shows the usefulness of the newfound bound of Eq.~(\ref{entbound}).  Indeed, it captures how local quantum correlations $E(\rho_{{\cal S}\varepsilon_i})$ are monotonically suppressed  by increasing the strength of the interaction between system and environment ($a\rightarrow 0$), while the known entropic limit monotonically increases. Also, it signals that quantum correlations (in fact, both quantum discord and entanglement) cannot exist without classical ones, as we will demonstrate in the next Section.
  


\section{Extending and clarifying limits to quantum information propagation}\label{sec3}

\subsection{Behavior of quantum discord in the proximity of classical states}

\noindent Here, we derive new results that show how bipartite quantum correlations are  restricted in many-body systems. We observe that the findings of the previous section imply that, if $\delta=0$, and therefore $J\left(\rho_{{\cal S}\check{\varepsilon}_{i}}\right)=H(\rho_{\cal S}),\,\forall\,i$, then there is no environment fragment that can share quantum discord with ${\cal S}$. We prove a statement about the degenerate case of this scenario:  all the subsystems store the very same amount of classical information about ${\cal S}$, but its value is zero, i.e.,  no classical correlations exist.\\
{\bf Remark:} {\it There are not quantum correlations without classical correlations:}
\begin{eqnarray}
J\left(\rho_{{\cal S}\check{\varepsilon}_{i}}\right)=0\Rightarrow D\left(\rho_{{\cal S}\check{\varepsilon}_{i}}\right) =0.
\end{eqnarray}
{\it Proof --} This claim can be proved in several ways. For example, from the Koashi-Winter relation, it follows that $J\left(\rho_{{\cal S}\check{\varepsilon}_{i}}\right)=0\Rightarrow E(\rho_{{\cal SE}_{/i}})=D\left(\rho_{{\cal S}\check{\cal E}_{/i}}\right)= H({\cal S})$. Since $E(\rho_{{\cal SE}_{/i}})+E\left(\rho_{{\cal S}{\varepsilon}_{i}}\right)= D\left(\rho_{{\cal S}\check{\varepsilon}_{i}}\right) + D(\rho_{{\cal S}\check{\cal E}_{/i}})$ \cite{fanchini}, one has $D\left(\rho_{{\cal S}\check{\varepsilon}_{i}}\right) =0$.\\

\noindent Next, we explore more nuanced aspects of the transition from quantum to classical regimes. We ask whether quantum discord is ``continuous'', in the sense of taking small values for states that are geometrically close (and physically similar) to classically correlated density matrices. The bound in  Eq.~(\ref{main}) establishes  that simultaneous maximal classical correlations between ${\cal S}$ and each $\varepsilon_i$ destroy  quantum discord  throughout the Universe. Hence, quantum information about ${\cal S}$ is not accessible to independent  observers that monitor different $\varepsilon_i$. 
Proving that quantum discord is subject to sharp continuity bounds  at the frontier with classical states would mean that, whenever a classical description of the correlation pattern is sufficiently precise, quantum correlations are inevitably negligible. That is, classical objectivity and a  significant amount of quantum correlations cannot co-exist.\\
It is known that $D\left(\rho_{{\cal S}\check{\cal F}_k}\right)=0$ if and only if there exists a measurement $\mathbf{M}_k$ such that $\rho_{{\cal SF}_{k}}=\rho_{{\cal SF}_{k,\mathbf{M}_k}}$. We  here prove  continuity bounds to quantum discord about the zero value.
 First we show that if a state of a partition ${\cal SF}_k$ (which we assume to be a full rank density matrix)  is close to the set of post-measurement states $\rho_{{\cal SF}_{k, \mathbf{M}_k}}$, then its discord is small. Given the subset of the projective measurements $\{\mathbf{P}_k\}\subset \{\mathbf{M}_k\}$ which can be performed on ${\cal F}_k$, recalling the definition of relative entropy $H\left(\rho_{\cal X}||\rho_{{\cal Y}}\right):=\text{Tr}\{\rho_{\cal X}\log_2\rho_{\cal X}\}-\text{Tr}\{\rho_{\cal X}\log_2\rho_{\cal Y}\}$, one has
\begin{align}
D\left(\rho_{{\cal S}\check{\cal F}_k}\right) &\leq \min_{\mathbf{P}_k}\left\{ I\left(\rho_{{\cal S}{\cal F}_k}\right)-J\left(\rho_{{\cal S}\check{\cal F}_k}\right)\right\}\\
 &= \min_{\mathbf{P}_k}\left\{H\left(\rho_{{\cal S}{\cal F}_k}||\rho_{{\cal S}}\otimes \rho_{{\cal F}_k}\right) -H\left(\rho_{{\cal S}{\cal F}_{k,\mathbf{P}_k}}||\rho_{{\cal S}}\otimes \rho_{{\cal F}_{k,\mathbf{P}_k}}\right) \right\}\nonumber\\ 
  &= \min_{\mathbf{P}_k}\left\{ H\left(\rho_{{\cal S}{\cal F}_k}||\rho_{{\cal S}{\cal F}_{k,\mathbf{P}_k}}\right) -H\left(\rho_{{\cal F}_k}||\rho_{{\cal F}_{k,\mathbf{P}_k}}\right) \right\}\nonumber\\
   &\leq \min_{\mathbf{P}_k}\left\{ H\left(\rho_{{\cal S}{\cal F}_k}||\rho_{{\cal SF}_{k,\mathbf{P}_k}}\right) \right\}.\nonumber
\end{align}
Finally, we obtain 
\begin{equation}\label{cont1}
 \min_{\mathbf{P}_k}\left\{ H\left(\rho_{{\cal S}{\cal F}_k}||\rho_{{\cal SF}_{k,\mathbf{P}_k}}\right) \right\} \leq \epsilon \Rightarrow D\left(\rho_{{\cal S}\check{\cal F}_k}\right) \leq \epsilon, \forall\,\epsilon.
\end{equation}
Therefore, states that are geometrically close ($\epsilon\rightarrow 0$) to be embeddings of classical probability distributions (classical-quantum states) display small values of quantum discord. \\ 
\noindent For the sake of completeness, we calculate the maximal relative entropy between a state and the closest classically correlated state when an upper bound to quantum discord, which we  obtain by maximizing in Eq.~(\ref{class}) over projective measurements rather than POVMs, takes arbitrary small values. As a preliminary step, we recall an upper limit to the relative entropy between two arbitrary states \cite{jens}:
\begin{align}
H\left(\rho_{{\cal X}}||\rho_{{\cal Y}} \right)&\leq (\lambda_{min}(\rho_{{\cal Y}})+d_{{\cal X},{\cal Y}})\, \log\left(1+\frac{d_{{\cal X},{\cal Y}}}{\lambda_{min}(\rho_{{\cal Y}})}\right) \\
&-\lambda_{min}(\rho_{{\cal X}})\,\log\left(1+\frac{d_{{\cal X},{\cal Y}}}{\lambda_{min}(\rho_{{\cal X}})}\right),\nonumber\\
d_{{\cal X},{\cal Y}}&\equiv||\rho_{{\cal X}}-\rho_{{\cal Y}}||_1/2,\nonumber
\end{align}
in which $\lambda_{min}(\rho_{{\cal X}})$ is the smallest eigenvalue of $\rho_{{\cal X}}$.
Then, calling $\tilde{\mathbf{P}}_k$ the projective measurement performed on ${\cal F}_k$ that maximizes the post-measurement mutual information (see Eq.~(\ref{class})), we obtain
\begin{align}
 H\left(\rho_{{\cal S}{\cal F}_k}||\rho_{{\cal S}{\cal F}_{k,\tilde{\mathbf{P}}_k}}\right) -H\left(\rho_{{\cal F}_k}||\rho_{{\cal F}_{k,\tilde{\mathbf{P}}_k}}\right)  &\leq \epsilon\Rightarrow\nonumber\\
  H\left(\rho_{{\cal S}{\cal F}_k}||\rho_{{\cal S}{\cal F}_{k,\tilde{\mathbf{P}}_k}}\right)  &\leq \epsilon +H\left(\rho_{{\cal F}_k}||\rho_{{\cal F}_{k,\tilde{\mathbf{P}}_k}}\right)\nonumber\\
    H\left(\rho_{{\cal S}{\cal F}_k}||\rho_{{\cal S}{\cal F}_{k,\tilde{\mathbf{P}}_k}}\right)  &\leq \epsilon + f\left(\rho_{{\cal F}_k}, \tilde{\mathbf{P}}_k\right),
\end{align}
where 
\begin{align*}\label{cont2}
  &f\left(\rho_{{\cal F}_k}, \tilde{\mathbf{P}}_k\right) =\\
   &\left\{\lambda_{min}\left(\rho_{{\cal F}_{k,\tilde{\mathbf{P}}_k}}\right)+d_{{\cal F}_k,{\cal F}_{k,\tilde{\mathbf{P}}_k}}\right\}\, \log\left\{1+\frac{d_{{\cal F}_k,{\cal F}_{k,\tilde{\mathbf{P}}_k}}}{\lambda_{min}\left(\rho_{{\cal F}_{k,\tilde{\mathbf{P}}_k}}\right)}\right\}\\
   &-\lambda_{min}\left(\rho_{{\cal F}_k}\right)\,\log\left(1+\frac{d_{{\cal F}_k,{\cal F}_{k,\tilde{\mathbf{P}}_k}}}{\lambda_{min}(\rho_{{\cal F}_k})}\right).
\end{align*}
This constraint is certainly less neat than Eq.~(\ref{cont1}) for generic mixed states. We leave to future studies to shape this claim, as we conjecture that a cleaner continuity bound may exist.
 
\subsection{Generalized bound to the entanglement of formation}
\noindent We now generalize the bound to the bipartite entanglement  in a star-like configuration (Eq.~(\ref{entbound})). We focus on the correlation structure of the environment ${\cal E}$, which is a generic $N$-partite quantum system.
We show that there is an upper bound to the  bipartite entanglement of formation between two components of the environment, in terms of how much classical information is shared by the environment parts. \\
We define a new disagreement quantifier: 
\begin{align}
\delta^\varepsilon_i:=1- \frac{\min\limits_{\varepsilon_j}
J\left(\rho_{\varepsilon_i\check\varepsilon_j}\right)}{H(\rho_{\varepsilon_i})}\, \in\,[0,1].
\end{align}
\noindent The quantity manifestly enjoys the same properties of the parameter   introduced in Eq.~(\ref{delta}).
Then, the entanglement of formation between an environment subsystem $\varepsilon_i$ and any other subsystem is limited by the (lack of) consensus about measurement outcomes  (classical information) on $\varepsilon_i$ across the environment. By employing again the Koashi-Winter relation, one has
\begin{align}\label{entbound2} 
J\left(\rho_{\varepsilon_i\check\varepsilon_j}\right)\geq&\,  (1-\delta^\varepsilon_i)\,H(\rho_{\varepsilon_i}),\,\forall\,i,j\Rightarrow \nonumber\\
E\left(\rho_{\varepsilon_{i}{\cal E}_{/ij}}\right)\leq&\, \delta^\varepsilon_i\,H(\rho_{\varepsilon_i}),\,\forall\,i,j \Rightarrow\nonumber\\
E\left(\rho_{\varepsilon_i\varepsilon_j}\right)\leq&\, \delta^\epsilon_i\,H(\rho_{\varepsilon_i}),\,\forall\, i,j.
\end{align}
The bound is clearly saturated, for example, for the GHZ state.
 
\section*{Conclusion} 
\noindent We have investigated quantitative limits to the propagation of quantum information in many-body systems. Specifically, we have extended the results of  \cite{red}, calculating a continuity bound to quantum discord nearby classical states (Eq.~(\ref{cont1})), and proving an upper bound to the entanglement of formation (Eq.~(\ref{entbound2})) between two arbitrary components of a multipartite system. \\
Classical correlations are not subject to any limitations. Consequently,  classical information can be freely broadcast from a source to an arbitrary number of receivers. Yet, the very same possibility that observers reach consensus on such classical information of target  physical systems dictates bounds to quantum information, which are here formulated in terms of limits to quantum discord and the entanglement of formation. \\
The results further corroborate the key ideas of Quantum Darwinism, a theoretical framework that explain the emergence of classical reality within quantum mechanics. We hope these findings will propel further studies on the subtleties of the transition between the quantum and classical regimes, which may lead to derive stronger bounds than the ones here demonstrated. Also, quantitative limits to genuinely multipartite quantum correlations may exist \cite{multi,cover}.
\begin{acknowledgments}
\noindent This research was supported by the Italian Ministry of Research, grant number MUR-PINR2022, Contract Number NEThEQS (2022B9P8LN). 
\end{acknowledgments}


\begin{thebibliography}{99}


\bibitem{epr2}A. Einstein, B. Podolsky, and N. Rosen, Can Quantum-Mechanical Description of Physical Reality be Considered Complete?", Phys. Rev. 47, 777–780 (1935)
\bibitem{EPR}N. Bohr, 
Can Quantum-Mechanical Description of Physical Reality be Considered Complete?, Phys. Rev. 48, 696 (1935).



\bibitem{entanglement}R. Horodecki, P. Horodecki, M. Horodecki, and K. Horodecki, Quantum entanglement, 
Rev. Mod. Phys. 81, 865 (2009).



\bibitem{telep}C. H. Bennett, G. Brassard, C. Crépeau, R. Jozsa, A. Peres, and W. K. Wootters, Teleporting an unknown quantum state via dual classical and Einstein-Podolsky-Rosen channels, Phys. Rev. Lett. 70, 1895 (1993).
\bibitem{super}C.  Bennett and S. Wiesner, Communication via one- and two-particle operators on Einstein-Podolsky-Rosen states,  Phys. Rev. Lett. 69, 2881 (1992).
\bibitem{capa} C. H. Bennett, P. W. Shor, J. A. Smolin, and A. V. Thapliyal, Entanglement-assisted classical capacity of noisy quantum
channel, Phys. Rev. Lett. 83, 3081 (1999).
\bibitem{naturedisc}
B. Dakić, Y. O. Lipp, X. Ma, M. Ringbauer, S. Kropatschek, S. Barz, T. Paterek, V. Vedral, A. Zeilinger, Č. Brukner, and P. Walther, Quantum discord as resource for remote state preparation,
Nature Phys. 8, 666 (2012).

\bibitem{tech}{\it Lectures on General Quantum Correlations and their Applications}, Eds.: F. F. Fanchini, D. Soares Pinto,  and G. Adesso, Springer (2017).


\bibitem{noclone}
W. K. Wootters and W. H. Zurek, A single quantum cannot be cloned,
Nature 299, 802 (1982).
\bibitem{mono}R. Prabhu, A. K. Pati, A. S. De, and U. Sen, Conditions for Monogamy of Quantum Discord: Greenberger-Horne-Zeilinger versus W states,
Phys. Rev. A 85, 040102(R) (2012).


\bibitem{brandao}
F. G. S. L. Brandao, M. Piani, and P. Horodecki, Generic emergence of classical features in quantum Darwinism, Nature Comm. 6, 7908 (2015).

\bibitem{adesso}P. A. Knott, T. Tufarelli, M. Piani, and G. Adesso, Generic Emergence of Objectivity of Observables in Infinite Dimensions, Phys. Rev. Lett. 121, 160401 (2018).


\bibitem{ranard}X.-L. Qi and D. Ranard, Emergent classicality in general multipartite states and channels, Quantum 5, 555 (2021).


\bibitem{akram}A. Touil, B. Yan, D. Girolami, S. Deffner, and W. H. Zurek, Eavesdropping on the Decohering Environment: Quantum Darwinism, Amplification, and the Origin of Objective Classical Reality, Phys. Rev. Lett. 128 (1), 010401 (2022).

\bibitem{akram2}
A. Touil, F. Anz\`a, S. Deffner, and J. P. Crutchfield, Branching States as The Emergent Structure of a Quantum Universe, arXiv:2208.05497.


\bibitem{red}
D. Girolami, A. Touil, B. Yan, S. Deffner, and W. H. Zurek, Redundantly amplified information suppresses quantum correlations in many-body systems,
Phys. Rev. Lett. 129, 010401 (2022).

\bibitem{darwin1}W. H. Zurek, Quantum Darwinism, Nature Phys. 5, 181 (2009).


\bibitem{darwin2}C. J. Riedel and W. H. Zurek, Quantum Darwinism in an Everyday Environment: Huge Redundancy in Scattered Photons, Phys. Rev. Lett. 105, 020404 (2010)

\bibitem{darwin4}J. K. Korbicz, R. Horodecki, and P. Horodecki, Objectivity Through State Broadcasting: The Origins Of Quantum Darwinism, Phys. Rev. Lett. 112, 120402 (2014).
 

\bibitem{darwin5}M. Zwolak and W. H. Zurek,  Complementarity of quantum discord and classically accessible information, Sci Rep 3, 1729 (2013).


\bibitem{darwin6}T. P. Le and A. Olaya-Castro, Strong Quantum Darwinism and Strong Independence is equivalent to Spectrum Broadcast Structure, Phys. Rev. Lett. 122, 010403 (2019).





\bibitem{exp2}T. Unden, D. Louzon, M. Zwolak, W. H. Zurek, and F. Jelezko, Revealing the Emergence of Classicality Using Nitrogen-Vacancy Centers, Phys. Rev. Lett. 123, 140402 (2019). 

\bibitem{pan}M.-C. Chen, H.-S. Zhong, Y. Li, D. Wu, X.-L. Wang, L. Li, N.-L. Liu, C.-Y. Lu, and J.-W. Pan, Emergence of Classical Objectivity of Quantum Darwinism in a Photonic Quantum Simulator, Science Bull. 64, 580 (2019).

\bibitem{ollivier2}H. Ollivier, D. Poulin, and W. H. Zurek, Environment as a witness: Selective proliferation of information and emergence of objectivity in a quantum universe, Phys. Rev. A 72, 042113 (2005).


\bibitem{korbiczreview}J. K. Korbicz, Roads to objectivity: Quantum Darwinism, Spectrum Broadcast Structures, and Strong quantum Darwinism – a review, Quantum 5, 571 (2021).
\bibitem{deco}W. H. Zurek, Decoherence, einselection, and the quantum origins of the classical, Rev. Mod. Phys. 75, 715 (2003).



\bibitem{Z07}W. H. Zurek, Quantum origin of quantum jumps: Breaking of unitary symmetry induced by information transfer and the transition from quantum to classical, 
Phys. Rev. A 76, 052110 (2007).


\bibitem{Z13}W. H. Zurek, Wave-packet collapse and the core quantum postulates: Discreteness of quantum jumps from unitarity, repeatability, 
and actionable information, Phys. Rev. A 87, 052111 (2013).


\bibitem{Zurek81}W. H. Zurek, Pointer basis of quantum apparatus: Into what mixture does the wave packet collapse?, Phys. Rev. D 24, 1516 (1981).


\bibitem{discordzurek}H. Ollivier and W. H. Zurek, Quantum Discord: A Measure of the Quantumness of Correlations, Phys. Rev. Lett. 88, 017901 (2001).


\bibitem{entform} C. H. Bennett, D. P. DiVincenzo, J. Smolin, and W. K. Wootters, Mixed-state entanglement and quantum error correction, 
Phys. Rev. A 54, 3824 (1996).


\bibitem{discorev}
K. Modi, A. Brodutch, H. Cable, T. Paterek, and V. Vedral, The classical-quantum boundary for correlations: discord and related measures, Rev. Mod. Phys. 84, 1655-1707 (2012).


 

\bibitem{corr1}Y. Huang, Computing quantum discord is NP-complete, New J. of Phys. 16, 033027 (2014).

\bibitem{corr2}S. Gharibian, Strong NP-hardness of the quantum separability problem, Quantum Inf. Comput.10, 343 (2010).

\bibitem{sanpera}G. De Chiara and A. Sanpera, Genuine quantum correlations in quantum many-body systems: a review of recent progress, Report on Prog. Phys. 81 074002 (2018).

\bibitem{newzurek}W. H. Zurek, Einselection and Decoherence from an Information Theory Perspective, Proceedings of the Planck constant centenary meeting,
 Ann. der Phys. (Leipzig) 9, 855 (2000).

\bibitem{sen}A. Bera, T. Das, D. Sadhukhan, S. Singha Roy, A. Sen De, and U. Sen, Quantum discord and its allies: a review,
Rep. Prog. Phys. 81 024001 (2018)


\bibitem{discome}D. Girolami and G. Adesso, Quantum discord for general two-qubit states: analytical progress, Phys. Rev. A 83, 052108 (2011).

\bibitem{koashi}M. Koashi and A. Winter, Monogamy of quantum entanglement and other correlations, Phys.
Rev. A 69, 022309 (2004).

\bibitem{braga}H. C. Braga, C. C. Rulli, Thiago R. de Oliveira, and M. S. Sarandy, Monogamy of Quantum Discord by Multipartite Correlations, Phys. Rev. A 86, 062106 (2012).

\bibitem{monogamy}
A. Streltsov, G. Adesso, M. Piani,  and D. Bruss, Are general quantum correlations monogamous?, Phys. Rev. Lett. 109, 050503 (2012).


\bibitem{acin}A. Ferraro, L. Aolita, D. Cavalcanti, F. M. Cucchietti, and A. Acín, Almost all quantum states have nonclassical correlations, Phys. Rev. A 81, 052318 (2010).


 
\bibitem{vedral}L. Henderson and V. Vedral, Classical, quantum and total correlations, J. of Phys. A: Math. and Gen. 34, 6899 (2001). 


\bibitem{holevo}A. S. Holevo, Bounds for the quantity of information transmitted by a quantum communication channel, Prob. of Inf. Trans. 9, 177 (1973).


\bibitem{wiseman} H. M. Wiseman, Quantum discord is Bohr's notion of non-mechanical disturbance introduced in his answer to EPR,  Ann. Phys. 338, 361 (2013).


\bibitem{streltsovzurek}A. Streltsov and W. H. Zurek, Quantum discord cannot be shared, Phys. Rev. Lett. 111, 040401 (2013).



\bibitem{locc}E. Chitambar, D. Leung, L. Mancinska, M. Ozols, and A. Winter, Everything You Always Wanted to Know About LOCC (But Were Afraid to Ask), Commun. Math. Phys.  328, 303 (2014).

 \bibitem{convert}J. Ma, B. Yadin, D. Girolami, V. Vedral, and M.  Gu, Converting Coherence to Quantum Correlations,
Phys. Rev. Lett. 116, 160407 (2016).

\bibitem{StreltsovBruss}A. Streltsov, H. Kampermann, and D. Bruss, Linking quantum discord to entanglement in a measurement, Phys. Rev. Lett. 106, 160401 (2011).

\bibitem{adessopiani}M. Piani, S. Gharibian, G. Adesso, J. Calsamiglia, P. Horodecki and A. Winter, All nonclassical correlations can be activated into distillable entanglement, Phys. Rev. Lett. 106, 220403 (2011).

\bibitem{lqu}D. Girolami, T. Tufarelli, and G. Adesso, Characterizing nonclassical correlations via local quantum uncertainty, Phys. Rev. Lett. 110, 240402 (2013).
\bibitem{fanchini}F. F. Fanchini, M. F. Cornelio, M. C. de Oliveira, and A. O. Caldeira, Conservation law for distributed entanglement of formation and quantum discord, Phys. Rev. A 84, 012313 (2011).




\bibitem{photon2}C. J. Riedel and W. H. Zurek, Redundant information from thermal illumination: quantum darwinism in scattered photons, New
J. Phys. 13, 073038 (2011).
 

\bibitem{ranktwo}M. Shi, W. Yang, F.  Jiang, and  J. Du, Quantum discord of two-qubit rank-two states,  J. Phys. A 44, 415304 (2011).

  \bibitem{jens}
K. M.R. Audenaert and J. Eisert, Continuity bounds on the quantum relative entropy - II, J. Math. Phys. 52, 112201 (2011).
 
 \bibitem{multi}D. Girolami, T. Tufarelli, and C. E.  Susa, Quantifying genuine multipartite correlations and their pattern complexity, Phys. Rev. Lett. 119, 140505 (2017).

\bibitem{cover}T. Cover and J. Thomas, {\it Elements of Information Theory}, Wiley (1991).



\end{thebibliography}
\end{document}